\documentclass[twocolumn,showpacs,preprintnumbers,amsmath,amssymb,footinbib]{revtex4}

\usepackage{graphicx}
\usepackage{dcolumn}
\usepackage{bm}
\usepackage{epstopdf}
\usepackage{enumerate}
\usepackage{mathrsfs}

\begin{document}

\preprint{APS/123-QED}

\title{Crystal length effects on the angular spectrum of spontaneous parametric downconversion photon pairs}

\author{R. Ram\'irez-Alarc\'on$^{1,2}$, H. Cruz-Ram\'irez$^{1}$ and A.B. U'Ren$^{1}$}

\affiliation{$^1$Instituto de Ciencias Nucleares, Universidad Nacional
Aut\'onoma de M\'exico, apdo. postal 70-543, M\'exico 04510 D.F.}
\address{$^2$Divisi\'on de Ciencias e Ingenier\'ia, Universidad de
Guanajuato, Loma del Bosque No. 103 Col. Lomas del Campestre C.P
37150 A.Postal E-143 Le\'on, Guanajuato.}


\date{\today}

\newcommand{\epsfg}[2]{\centerline{\scalebox{#2}{\epsfbox{#1}}}}

\begin{abstract}
We present a theoretical and experimental analysis of the joint effects of the
transverse electric field distribution and of the nonlinear crystal characteristics
on the properties of photon pairs generated by spontaneous
parametric downconversion (SPDC).    While it is known that for a sufficiently short crystal
the pump electric field distribution fully determines the joint signal-idler properties,
for longer crystals the nonlinear crystal properties also play an important role.
In this paper we present experimental measurements of the angular spectrum (AS)
and of the conditional angular spectrum (CAS) of photon pairs
produced by spontaneous parametric downconversion (SPDC), carried
out through spatially-resolved photon counting.    In our experiment we control whether or not the
source operates in the short-crystal regime through the degree of pump focusing, and
explicitly show how the AS and CAS measurements differ in these two regimes.  Our
theory provides an understanding of the boundary between these two regimes and also
predicts the corresponding differing behaviors.
\end{abstract}

\maketitle



\section{Introduction}

The process of spontaneous parametric downconversion
(SPDC)~\cite{burnham70} has been used for the generation of paired
photons in many recent experiments, ranging from fundamental tests
of quantum mechanics~\cite{zeilinger99} to implementations of
quantum information processing protocols~\cite{kok07}.  In this
process a laser pump beam illuminates a second-order non-linear
crystal, where individual pump photons are annihilated giving rise
to the emission of signal and idler photon pairs.  In particular,
these photon pairs exhibit a rich transverse spatial
structure~\cite{rubin96,joobeur94}, which forms the subject matter
of the present study.  Photon pairs entangled in the spatial degree
of freedom, including the specific case of entanglement in orbital
angular momentum~\cite{mair01,molina-terriza07}, are of interest
because each photon can be described by a multi-dimensional Hilbert
space~\cite{howell04,law04,walborn07,fedorov08,straupe11,dilorenzopires09}
compared to the case of polarization which is limited to a dimension
of $2$~\cite{kwiat95}.


Spatially-resolved single-photon detection in the transverse
momentum domain yields the angular spectrum (AS) of the
SPDC bi-photon field, which for type-I SPDC has a well-known annular
shape.    If a single photon is detected at a certain
location in the AS with transverse momentum value $\tilde{\textbf{k}}^\bot$, the conjugate photon can be detected in
coincidence around the location which fulfills transverse momentum conservation, i.e. with transverse momentum value $-\tilde{\textbf{k}}^\bot$.  This forms the basis for
a second measurement of interest, in which the idler photon is detected
at a certain transverse momentum value and where spatially-resolved
coincidence detection of the signal photon yields the
\textit{conditional} angular spectrum (CAS), which may be thought of as
the shape of the heralded signal-mode photon~\cite{torres05}.
An idealized plane-wave pump in the process of SPDC leads to strict
transverse momentum correlations, so that each individual
idler-mode $k$-vector in the AS is correlated to a
single signal-mode $k$-vector, yielding a delta-like CAS. In contrast, for a realistic experimental
situation involving finite transverse dimensions of the pump beam,
these transverse momentum correlations are no longer strict, i.e.
each individual idler-mode $k$-vector is correlated with a
\emph{spread} of signal-mode $k$-vectors, so that the CAS  acquires a certain width.
%

The properties of spatially-entangled SPDC photon pairs, including
the AS and CAS, are determined on the one hand by the transverse
electric field distribution of the pump, and on the other hand by
the nonlinear crystal properties, including crystal dispersion,
length and Poynting vector walk off.  The manner in which the pump
transverse spatial properties are mapped to those of the SPDC photon
pairs has been studied in a number of
papers~\cite{monken98,walborn10,joobeur94,law04,warborn03,torres05,grayson94}.
It is known that for a sufficiently short crystal, the
spatially-resolved rate of signal-idler coincidences is fully
determined by the pump transverse electric field
distribution~\cite{monken98,walborn10,molina05,saleh00,lorenzo09,neves04bis,lee05,walborn07,jha10}.
Within this limit, it thus becomes straightforward to engineer
photon pairs to have a particular spatial character, as determined
by the pump.  This forms  the basis for a large body of work, which
seeks to exploit particular types of spatial correlations present in
SPDC photon pairs. Indeed, the short-crystal approximation has been
used in the context of the implementation of quantum information
processing protocols, in experiments which exploit the orbital
angular momentum of SPDC photon
pairs~\cite{molina03,arnaut01,walborn04,peeters07,ren05,kawase09},
in ghost imaging and diffraction
experiments~\cite{pittman96,santos03,abouraddy01,almeida06,costa10,walborn11},
as well as in a variety of other recent
experiments~\cite{fonseca99,fonseca99bis,walborn05,monken98bis,fonseca99ter,shimizu03,vidal08,neves06,lee06,deng06,ether06}.
However, for a sufficiently long crystal or for a sufficiently
focused pump beam, the photon pair properties are no longer solely
determined by the pump spatial distribution.   The objective of this
paper is to provide an understanding of the SPDC photon-pair
properties in general, including situations for which the
short-crystal approximation can and cannot be used, as well as of
the boundary between these two regimes.

We have performed a detailed study of the CAS and of how the AS properties are derived from those of the CAS,
paying special attention to crystal length effects.
We show that the CAS is given by the product of two separate functions, one which is related to the pump AS, and
another one which is related to the properties of the nonlinear crystal, including length, dispersion and Poynting vector walkoff.
In our study of crystal length effects, we show that a critical length $L_c$ exists, which depends on the degree of pump focusing, so that for
$L<L_c$, crystal length effects can be neglected, and the CAS is fully determined by the pump AS;  we show that  in this case the CAS is a displaced
version of the pump AS.  We also show that for $L>L_c$, the CAS is determined
both by the pump AS and the crystal properties.    In this latter case, we show that the CAS becomes elongated
to a degree related to the crystal length and tilted according to the location of the fixed detector, leading to azimuthal distinguishability.

We have carried out experimental measurements of the AS and CAS,
exhibiting excellent agreement with numerical simulations based on
our SPDC theory, which is general enough to include essentially all
experimental aspects of interest, i.e an arbitrary pump spatial
distribution, spatial and spectral filtering of the SPDC photons,
crystal dispersion, Poynting vector walkoff, and the spatial extent
of detectors.    In particular, we have performed a careful
experimental/numerical exploration of the AS and of the CAS, where
we vary the degree of pump
focusing~\cite{klyshko82,lee05,bennink06,dilorenzopires11,suzer08,grice11,vicent10}
in order to explore and contrast the resulting behavior in the
$L<L_c$ and $L>L_c$ regimes. Our theory presented here explains the
azimuthal distinguishability evident in the CAS measurements of
Ref.~\cite{osorio07}, making it clear that this distinguishability
is a crystal-length effect which appears only for sufficiently long
crystals ($L>L_c$).   This paper leads to a quantitative and
qualitative understanding, not available in previous works,  of how
the spatial character of SPDC photon pairs is determined by
nonlinear crystal properties in addition to the pump angular
spectrum, which should be useful for the design of photon pair
sources for specific applications.


\section{Theory}

The quantum state which describes photon pairs produced by SPDC is
given by $|\Psi\rangle=|\mbox{vac}\rangle+ \eta |\Psi_2\rangle$, with

\begin{eqnarray}\label{E:state}
\left|\Psi_2\right\rangle
    &&=
   \int d \omega_s \int d^2 \textbf{k}_s^\bot
    \int d \omega_i \int d^2 \textbf{k}_i^\bot
    F(\omega_s,\textbf{k}^\bot_s, \omega_i,\textbf{k}^\bot_i)
   \nonumber\\
    &&\times\hat{a}^\dag(\omega_s,\textbf{k}^\bot_s)
    \hat{a}^\dag(\omega_i,\textbf{k}^\bot_i)
    \left|\mbox{vac}\right\rangle
\end{eqnarray}

\noindent where $\eta$ is a constant related to the conversion
efficiency, $F(\omega_s,\textbf{k}^\bot_s, \omega_i,\textbf{k}^\bot_i)$
represents the joint amplitude,
$\hat{a}^\dag_\mu(\omega_s,\textbf{k}^\bot_\mu)$
 (with $\mu=s,i$) is the creation
operator associated with the signal and idler modes, expressed
as a function of the transverse wavevector $\textbf{k}^\bot_\mu$ and frequency $\omega_\mu$,
and where
$|\mbox{vac}\rangle$ is the vacuum. Taking into account spectral filters
$f_\mu(\omega)$ applied to the signal and idler modes ($\mu=s,i$),
the joint amplitude  is given by

\begin{eqnarray}\label{E:JWA}
&F(\omega_s,\textbf{k}^\bot_s, \omega_i,\textbf{k}^\bot_i) =  A_s  \ell(\omega_s) A_i
   \ell(\omega_i) \,f_s(\omega_s)f_i(\omega_i) \nonumber \\
  &\times \phi(\omega_s,\textbf{k}^\bot_s, \omega_i,\textbf{k}^\bot_i) \alpha(\omega_s+\omega_i).
  \end{eqnarray}

In Eq.~\ref{E:JWA}, we have used the definitions $\ell(\omega) \equiv
\sqrt{\hbar\,\omega/[2(2\pi)^3 \epsilon_0\,n(\omega)^2]}$, where
$\epsilon_0$ is the permittivity of free space and $n(\omega)$ is
the index of refraction, and $A_\mu \equiv k_\mu' k_\mu/k_{\mu z}$.   $\phi(\omega_s,\textbf{k}^\bot_s, \omega_i,\textbf{k}^\bot_i)$ is the
phasematching function and $\alpha(\omega)$ is the spectral amplitude of
the pump.   $k_{\mu z}$ represents the longitudinal components of the signal and idler
$k$-vectors, given by  $k_{\mu z}=(k_\mu^2-|\textbf{k}_\mu^\bot|^2)^{1/2}$, with $k_\mu=n(\omega_\mu)\omega_\mu/c$.  Primed quantities denote frequencyt derivatives.  Let us define the transverse vector sum $\textbf{k}_{+}^\bot=\textbf{k}^\bot_s+ \textbf{k}^\bot_i$,
in terms of which the phase matching function can
be expressed as

\begin{eqnarray}\label{E:PMF}
\phi(\omega_s,\textbf{k}^\bot_s, \omega_i,\textbf{k}^\bot_i) &=&
S(\textbf{k}_{+}^\bot)  \mbox{sinc}\left(\frac{1}{2} L \Delta k(\omega_s,\textbf{k}^\bot_s, \omega_i,\textbf{k}^\bot_i) \right)  \nonumber\\
&\times& \mbox{exp}\left(i \frac{1}{2} L \Delta k(\omega_s,\textbf{k}^\bot_s, \omega_i,\textbf{k}^\bot_i) \right).
\end{eqnarray}

Here, the function $S(\textbf{k}^\bot)$ represents the pump
transverse wave vector amplitude distribution,
evaluated in the transverse wavevector sum $\textbf{k}_{+}^\bot$, so that $|S(\textbf{k}^\bot)|^2$ represents the pump AS.
The phasemismatch $\Delta k(\omega_s,\textbf{k}^\bot_s, \omega_i,\textbf{k}^\bot_i)$ can be expressed in
terms of the pump wavenumber $k_p$, the Poynting vector walkoff angle experienced by
the pump, $\rho_0$, and the y-component of the $\textbf{k}_{+}^\bot$ vector as

\begin{equation}
\Delta k(\omega_s,\textbf{k}^\bot_s, \omega_i,\textbf{k}^\bot_i) = k_p - \frac{|\textbf{k}_{+}^\bot|^2}{2\,k_p} - k_{sz} -k_{iz} -
k_{+y}^\bot\,\mbox{tan}\rho_0.
\end{equation}

Note that here we have assumed without loss of generality that walkoff occurs on the plane $zy$.
In this paper we are interested in studying the transverse spatial
structure of the emitted photon pairs,  specifically through
spatially-resolved photon counting experiments. Standard Fourier
optics techniques may be used in order to probe this structure in
either the transverse position, or the transverse wavevector
momentum domains.      Specifically, a map of
counts as a function of the signal- and idler-mode
transverse wavevector may be obtained by a detection scheme with
transverse spatial resolution on the Fourier plane located a
distance of one focal length $f$ from a lens of focal length
$f$, itself placed a distance $f$ from the SPDC crystal.
If two detectors are placed on the Fourier plane so that they collect photons with transverse
wavevectors $\textbf{k}^\bot_s$ and
$\textbf{k}^\bot_i$ and frequencies $\omega_s$ and $\omega_i$,  the  rate of signal and idler
coincidences is given by

\begin{eqnarray}\label{E:completeCAS}
&&R_c(\omega_s,\textbf{k}^\bot_s,\omega_i,\textbf{k}_i^\bot)  \nonumber \\
&&=  \langle \Psi_2
|a^\dag(\omega_s,\textbf{k}_s^\bot)a^\dag(\omega_i,\textbf{k}_i^\bot)a(\omega_i,\textbf{k}_i^\bot)a(\omega_s,\textbf{k}_s^\bot)
|\Psi_2 \rangle.
\end{eqnarray}

In a given experimental situation, this rate of detection should be integrated over the transverse wavevector and spectral acceptance of the detectors.

We will specialize our discussion to the case of
a continuous-wave pump, for which the pump may be regarded as
essentially monochromatic at frequency $\omega_p$, and
$|\alpha(\omega)|^2$ may be replaced by $\delta(\omega-\omega_p)$.
Assuming detectors with ideal transverse wavevector resolution, and integrating over
the spectral content of the photon pairs,  it may be shown that the rate of coincidences can then be written as follows

\begin{eqnarray}\label{E:CAS:PR}
R^{(0)}_c(\textbf{k}^\bot_s,\textbf{k}_i^\bot)=| S(\textbf{k}_{+}^\bot)|^2 \mathscr{L}(\textbf{k}^\bot_s,\textbf{k}^\bot_i).
\end{eqnarray}

Thus,  the rate of coincidences can be
factored into two contributions.  On the one hand, $|S(\textbf{k}_{+}^\bot)|^2$ is related to transverse phasematching and is
fully determined by the pump angular spectrum.    On the other hand,
$\mathscr{L}(\textbf{k}^\bot_s,\textbf{k}^\bot_i)$ is related to longitudinal phasematching and is determined by  the crystal, including the effects of  the crystal length, dispersion and Poynting vector walkoff.  This function is given by

\begin{eqnarray}\label{E:cursL}
\mathscr{L}(\textbf{k}_{s}^\bot,\textbf{k}_{i}^\bot)
    && =\int\mbox{d}\omega_i\,
    \frac{k_s' k_s}{k_{sz}}  \frac{k_i'k_i}{k_{iz}}
    |f(\omega_p-\omega_i)|^2\,|f(\omega_i)|^2 \nonumber\\
    &&\times   \mbox{sinc}^2\left[\frac{1}{2}\,L\,\Delta k(\omega_p-\omega_i,\textbf{k}^\bot_s, \omega_i,\textbf{k}^\bot_i)\right].
\end{eqnarray}

Note that while $k_\mu$ and $k'_\mu$ (with $\mu=s,i$) are functions
of the signal and idler frequencies, $\omega_p-\omega_i$ and $\omega_i$,
$k_{\mu z}$ is a function of these frequencies and of the corresponding transverse wavevector components.  Let
us consider the limit in which the pump beam is in the form of a
plane wave,  with transverse wavevector $\textbf{k}^\bot_{p}=0$.  In
this case, the function $|S(\textbf{k}^\bot_{+})|^2$ becomes

\begin{equation}
|S(\textbf{k}^\bot_{+})|^2=\delta(\textbf{k}^\bot_{+})=\delta(\textbf{k}^\bot_{s}+\textbf{k}^\bot_{i})
\end{equation}

This equation tells us that if a single idler photon is detected at
$\textbf{k}^\bot_{i}=\textbf{k}^\bot_{i0}$, the conjugate photons
may be found at $\textbf{k}^\bot_{s}=-\textbf{k}^\bot_{i0}$ so that
transverse momentum is exactly conserved.  Note, however, that the
probability of observing photon pairs at two such conjugate points
is limited by the function
$\mathscr{L}(-\textbf{k}_{i0}^\bot,\textbf{k}_{i0}^\bot)$, i.e. by
the existence of longitudinal phasematching at these two transverse
wavevector values.

Let us now consider the more general case where the pump is given by
a superposition of plane waves, i.e. for which the pump AS is no
longer a delta function.  In this case, if a single photon
is detected at  $\textbf{k}^\bot_{i}=\textbf{k}^\bot_{i0}$
on the transverse wavevector space, the conjugate photons may be
found \emph{around}  $\textbf{k}^\bot_{s}= -
\textbf{k}^\bot_{i0}$ with an uncertainty which grows with the width
of the pump angular spectrum.    We then refer to the function
$R_c(\textbf{k}^\bot_{s},\textbf{k}^\bot_{i0})$, which determines
this uncertainty, as the conditional angular spectrum (CAS) of the
signal photon, conditioned on the detection of a single idler photon
with transverse wavevector $\textbf{k}^\bot_{i0}$.  The function
$|S(\textbf{k}^\bot_s+ \textbf{k}^\bot_{i0})|^2$, which for a
sufficiently broad
$\mathscr{L}(\textbf{k}_{s}^\bot,\textbf{k}_{i0}^\bot)$ represents
the signal-photon CAS,   is a displaced version of the
pump AS, centered at $\textbf{k}^\bot_s=-\textbf{k}^\bot_{i0}$.

Specifically, let us consider the case where the pump beam is in the
form of a Gaussian beam with widths $W_x$ and $W_y$ along
the $x$ and $y$ directions.   In this case, $|S(\textbf{k}^\bot_{+})|^2$ is given by

\begin{eqnarray}
|S(\textbf{k}^\bot_{+})|^2=\exp\biggl(-\frac{1}{2}\{W_x^2\,(k_{+x}^\bot)^2
+ W_y^2\,(k_{+y}^\bot)^2\}\biggr),
\end{eqnarray}

\noindent in terms of the $x$ and $y$ components of the vector
$\textbf{k}_{+}^\bot$. As the  pump beam is increasingly focused,
corresponding to smaller values of $W_x$ and $W_y$, the pump AS
becomes broader also leading to a broader CAS, as limited by the
function $\mathscr{L}(\textbf{k}_{s}^\bot,\textbf{k}_{i}^\bot)$.
Thus, the strict one-to-one transverse momentum signal and idler
correlations which appear in the plane-wave pump limit become weaker
as the pump is increasingly focused.

So far we have considered idealized detection of the signal and
idler modes involving a single transverse wavevector value. However,
in a realistic experimental implementation, the transverse
dimensions of the detectors used for the signal and idler modes are
non-vanishing.     Suppose that the transverse wavevector acceptance
of each detector is characterized by functions
$u_s(\textbf{k}^\bot-\textbf{k}^\bot_s)$ and
$u_i(\textbf{k}^\bot-\textbf{k}^\bot_i)$ for the signal and idler
modes respectively, where each of the detectors is centered at
$\textbf{k}^\bot=\textbf{k}^\bot_\mu$ (with $\mu=s,i$).  Then, the
resulting coincidence rate obtained with these detectors can be
written as

\begin{eqnarray}\label{E:CAS}
R_c(\textbf{k}^\bot_s,\textbf{k}^\bot_i)&=& \int d^2 \tilde{\textbf{k}}^\bot_s \int d^2 \tilde{\textbf{k}}^\bot_i
R^{(0)}_c(\tilde{\textbf{k}}^{\bot}_s,\tilde{\textbf{k}}^{\bot}_i) \nonumber \\
&\times& u_s(\tilde{\textbf{k}}^\bot_s-\textbf{k}^\bot_s)u_i(\tilde{\textbf{k}}^\bot_i-\textbf{k}^\bot_i).
\end{eqnarray}

Let us now turn our attention to single-channel counts, i.e. obtained through a single detector. The rate of single-channel counts
obtained by a detector placed so that it selects single photons with transverse wavevector $\textbf{k}^\bot_s$ and frequency $\omega_s$, is given by
\begin{eqnarray}\label{E:completeAS}
&R_s(\omega_s,\textbf{k}^\bot_s)
=
&\langle \Psi_2
|a^\dag(\omega_s,\textbf{k}_s^\bot)a(\omega_s,\textbf{k}_s^\bot)
|\Psi_2 \rangle
\end{eqnarray}

In a given experimental situation, this rate of detection should be integrated over the transverse wavevector and spectral acceptance of the detector.
It may be shown that under the same conditions in which Eq.~\ref{E:CAS:PR} was derived,  the rate of single-channel detection is related to the CAS through the following simple relationship

\begin{eqnarray}\label{E:AS:PR}
R_s^{(0)}(\textbf{k}^\bot_s) = \int d^2  \tilde{\textbf{k}}^\bot_i R^{(0)}_c(\textbf{k}^\bot_s,\tilde{\textbf{k}}_i^\bot).
\end{eqnarray}

The above quantity represents the transverse-wavevector distribution, or angular spectrum (AS),
 of the SPDC photon pairs.  Thus, according to Eq.~\ref{E:CAS:PR},  the SPDC AS evaluated at a wavector $\textbf{k}^\bot_{s}=\textbf{k}^\bot_{s0}$ is given
by the CAS $R^{(0)}_c(\textbf{k}^\bot_{s0},\textbf{k}_i^\bot)$ integrated over all $\textbf{k}_i^\bot$
values.

When evaluating Eq.~\ref{E:AS:PR}  for type-I non-collinear SPDC one obtains a
well-known annular structure on the $\textbf{k}_{s}^{\bot}$ plane.  In
the next section, we will show experimental results, as well as
simulations based on Eq.~\ref{E:AS:PR}, which show this annular
structure, and how it differs in the two regimes of interest, namely 
those for which the short-crystal approximation can and cannot be used.

In the case of a non-ideal detector characterized by an acceptance function
$u_s(\textbf{k}^\bot-\textbf{k}^\bot_s)$, the signal-mode angular spectrum may be written as

\begin{eqnarray}\label{E:AS}
R_s(\textbf{k}^\bot_s)= \int d^2 \tilde{\textbf{k}}^\bot_s \int d^2 \tilde{\textbf{k}}^\bot_i
R^{(0)}_c(\tilde{\textbf{k}}^{\bot}_s,\tilde{\textbf{k}}^{\bot}_i) u(\tilde{\textbf{k}}^\bot_s-\textbf{k}^\bot_s).
\end{eqnarray}

For the discussion of the experimental results, below, it is useful
to consider the single-channel and double-channel detection rates as a function of the transverse
coordinates on the Fourier plane, behind an $f$-$f$ optical system, i.e. $R_s(\bm{\rho}_{s}^{\bot})$ and $R_c(\bm{\rho}_{s}^{\bot},\bm{\rho}_{i}^{\bot})$.  If
the signal and idler modes each involve a single emission frequency,
then these functions  are simply scaled versions of their counterparts in the wavevector domain,
according to the transformation $\textbf{k}^\bot_\mu=[\omega
/(c\,f)] \bm{\rho}^\bot_\mu$  (with $\mu=s,i$), where $f$ is the focal length used in the $f$-$f$ optical system, i.e. each transverse position on the
Fourier plane corresponds to a specific transverse momentum value.
However, because this transformation is frequency-dependent, if the
emitted modes contain a spread of frequencies the rate of single-channel detection in the position and
wavevector domains are not  simply scaled versions of each other.   When carrying
out simulations, care must be taken to integrate the functions $R_s(\bm{\rho}_{s}^{\bot})$ and $R_c(\bm{\rho}_{s}^{\bot},\bm{\rho}_{i}^{\bot})$
over the detected spectral components.

\begin{figure}[ht]
\centering
\includegraphics[width=8cm]{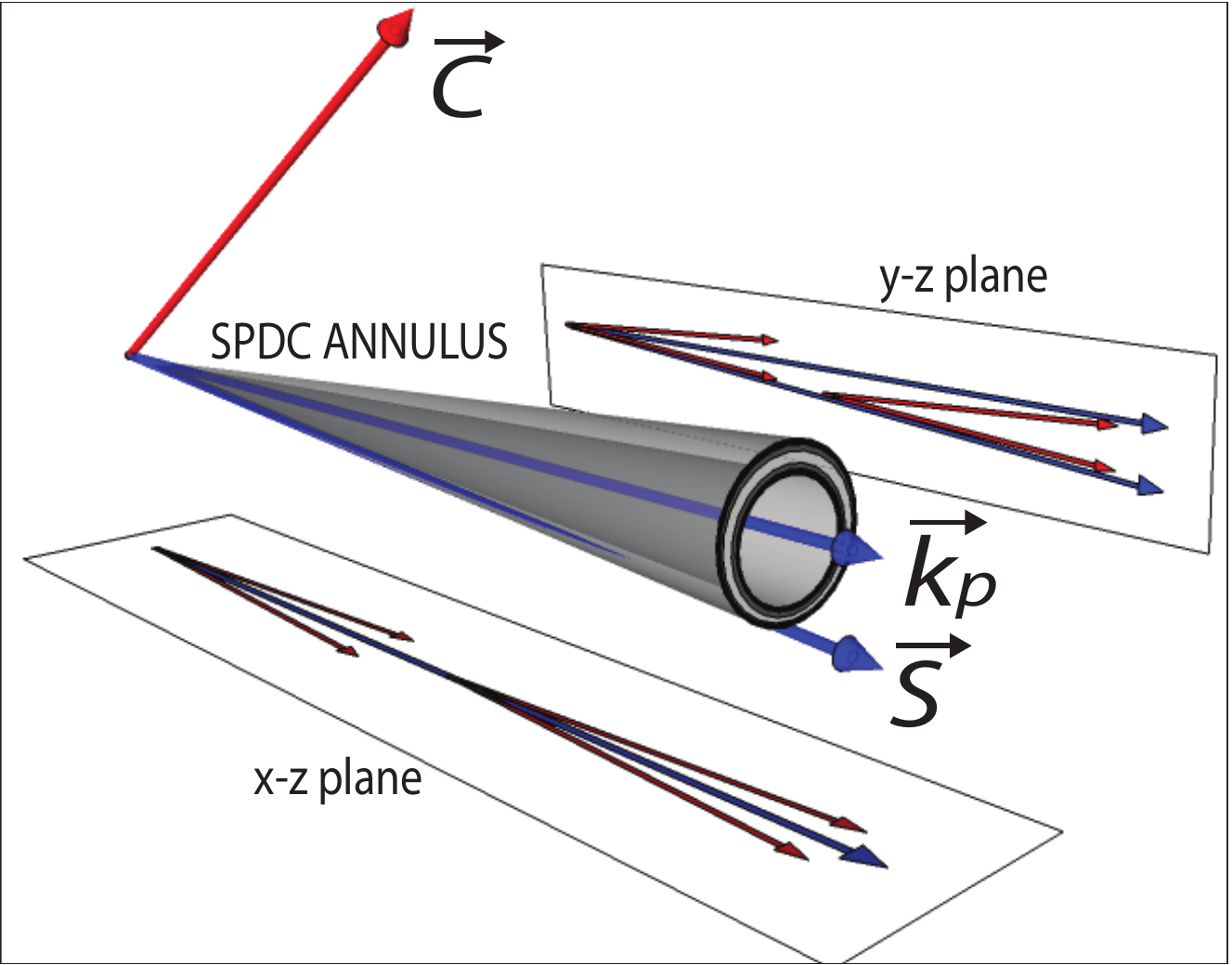}
\caption{(color online) Schematic of emission annulus with various  source
characteristics indicated.}\label{Fig:geometry}
\end{figure}

In Fig.~\ref{Fig:geometry} we show the geometry of the SPDC emission annulus
for non-collinear type-I SPDC from a negative uniaxial, specifically
beta barium borate (BBO), crystal.
We denote by $\vec{k}_p$ the wavevector
corresponding to the central direction of propagation of the pump beam,  by $\vec{S}$
the pump Poynting vector, and by $\vec{C}$ the crystal axis.  Note that the angular
separation between $\vec{S}$  and  $\vec{k}_p$ is due to Poynting vector walkoff. We also show
two pairs of signal and idler rays born at two distinct planes, projected
on the $y$-$z$ and $x$-$z$ planes.     As may be appreciated in the figure, because
photon pairs are born along the path of $\vec{S}$, on the $y$-$z$
plane  the two upper rays have a greater
separation from $\vec{S}$ compared to the two lower rays,
while the two corresponding separations are equal on the $x$-$z$
plane.  This leads to an asymmetry in the emission annulus.
Note that while for the above argument, we
have implicitly assumed a transversely well-localized pump beam, the pump
beam in fact has horizontal and vertical widths $W_x$ and $W_y$. If
these widths are considerably larger that the lateral ray
displacement $L \tan\rho_0$, the annulus asymmetry is in fact
suppressed.  In other words, this asymmetry is visible only for a
sufficiently focused beam so that $W_x,W_y \lesssim L \tan \rho_0$.
Note that this asymmetry, mediated by pump
focusing and Poynting vector walkoff, is the origin of so called
``hot spots'' which have been observed in the spatial flux
distribution  of type-II parametric downconversion~\cite{koch95}.

\begin{figure}[ht]
\centering
\includegraphics[width=8cm]{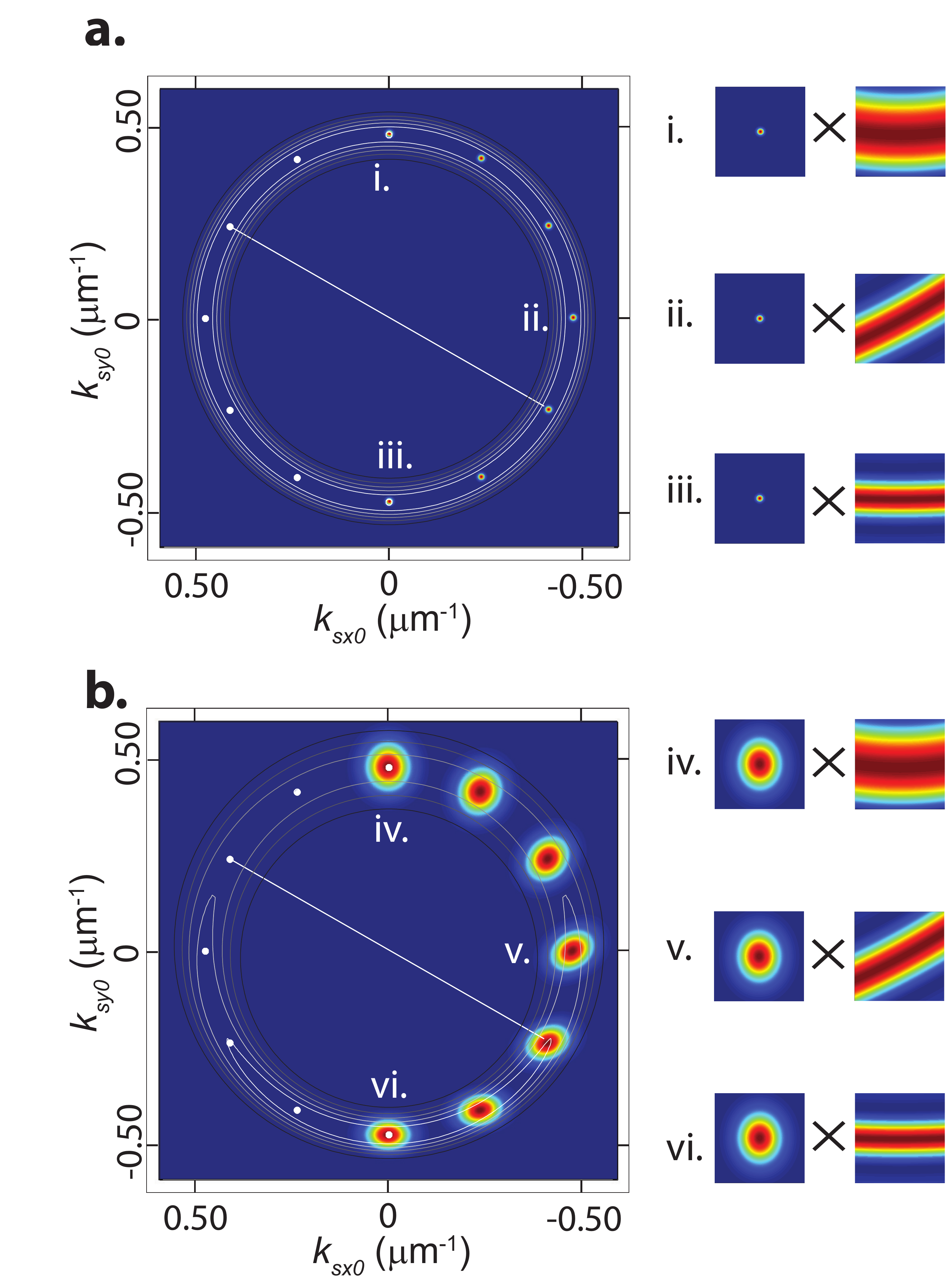}
\caption{(color online) Contour plot of the angular spectrum
for two different experimental situations (panel a: $W_x=182.0\mu$m, and $W_y=189.0 \mu$m and panel b: $W_x=38.9\mu$m, and $W_y=34.7 \mu$m), along with white dots covering one half of the
annulus,  angularly equi-spaced by $30^\circ$, along with the
conditional angular spectrum  corresponding to each of these white
dots.   For each of the two panels we also show, for the CAS corresponding to a fixed detector on the bottom, left and top of the AS, plots of the functions $|S(\textbf{k}^\bot_s+\textbf{k}^\bot_i)|^2$ and $\mathscr{L}(\textbf{k}^\bot_s,\textbf{k}^\bot_i)$ the product of which yields the CAS. }\label{Fig:fluxspdc}
\end{figure}

In Fig.~\ref{Fig:fluxspdc} we show simulations of the AS for the case of a Gaussian beam pump, based on numerical integration of Eqns.~\ref{E:CAS:PR}
and \ref{E:AS:PR}.  The AS is shown as a contour plot for two different degrees of focusing ($W_x=182.0\mu$m, $W_y=189.0 \mu$m for
panel a and $W_x=38.9\mu$m, $W_y=34.7 \mu$m for panel b; note that these choices of beam widths correspond to experimental situations presented below).
For these plots we have assumed a $1$mm long BBO crystal cut at $29.3^\circ$, for non collinear type-I phasematching.
Note that while the annulus is symmetric, with a constant width, in the
case of panel a, it becomes asymmetric with an azimuthally-varying width for panel b, as indeed is to be expected from the argument in the previous paragraph. This annulus asymmetry apparent in the single-channel counts spatial
distribution also translates into azimuthal distinguishability of
the photon pairs around the annulus (i.e. into an azimuthal variation of the orientation and width of the CAS)~\cite{osorio07}.    In order to illustrate this, in both panels of Fig.~\ref{Fig:fluxspdc} we also show the CAS
corresponding to $7$ points (shown as white dots) chosen to be angularly equidistant on the left-hand side of the angular spectrum.    For three of these points (top, left and bottom)
we show in addition plots of the $|S(\textbf{k}_{+}^\bot)|^2$ and $\mathscr{L}(\textbf{k}^\bot_s,\textbf{k}^\bot_i)$ functions, the product of which yields the CAS.  Note that the AS asymmetry may be explained in terms of clipping of the $|S(\textbf{k}_{+}^\bot)|^2$ function by the $\mathscr{L}(\textbf{k}^\bot_s,\textbf{k}^\bot_i)$ function.  For example, for a fixed detector at the top of the annulus, the corresponding CAS in a focused pump regime is narrowed by the horizontal structure of the $\mathscr{L}(\textbf{k}^\bot_s,\textbf{k}^\bot_i)$ function.  Thus, the AS at the top of the annulus given as the integral over all $\textbf{k}^\bot_i$ values of the CAS will have a lower value compared, say, to the diametrically opposed portion of the annulus where this clipping does not occur.  As will be discussed below, while panel $a$ corresponds to the short-crystal $L<L_c$ regime, panel $b$ corresponds to the $L>L_c$ regime.

It is interesting to relate the CAS in the spontaneous case which we study in this paper, to the size
of speckles obtained in the spatial intensity distribution in the
case of high-gain, i.e. stimulated  parametric
downconversion~\cite{brambilla04,
jedrkiewicz04,jedrkiewicz06,blanchet10,brida09,blanchet08}. Indeed,
SPDC photons with a given transverse angular momentum $\textbf{k}^\bot$
within the AS  can serve as a seed for parametric
amplification, leading to the appearance of coupled speckles at
$\textbf{k}^\bot$ and $-\textbf{k}^\bot$. According to the analysis in
Refs.~\cite{brambilla04,brida09}, the mean speckle area (also
sometimes referred to as coherence area), is inversely proportional
to the pump beam transverse area, an effect which mimics the
observed dependence of the area in transverse momentum space of our
CAS on the focusing strength.

Note on the one hand that the function  $\mathscr{L}(\textbf{k}^\bot_s,\textbf{k}^\bot_i)$  depends only on crystal properties; in particular, its width is determined by the crystal length $L$ with longer crystals yielding narrower widths.   Note on the other hand that the function $|S(\textbf{k}_{+}^\bot)|^2$ depends only on pump properties, and its width corresponds to the pump angular width, i.e. it is determined by the degree of pump focusing.   Thus, for a given degree of pump focusing there is a critical crystal length $L_c$ such that for $L <  L_c$ the function $\mathscr{L}(\textbf{k}^\bot_s,\textbf{k}^\bot_i)$  is wider than the function $|S(\textbf{k}_{+}^\bot)|^2$  so that the latter fully determines the CAS.  In contrast, for $L >L_c$, the CAS is determined by both of these functions together, i.e. by crystal properties in addition to pump properties.       Note from Fig.~\ref{Fig:fluxspdc} that a plot of the function $\mathscr{L}(\textbf{k}^\bot_s,\textbf{k}^\bot_i)$ yields a stripe which is horizontally-oriented at the top and bottom of the annulus, with a larger width at the top, and which is oriented diagonally at other annulus locations, with a maximum tilt at the left and right.  This azimuthal variability of the function  $\mathscr{L}(\textbf{k}^\bot_s,\textbf{k}^\bot_i)$ is the origin of the azimuthal distinguishability of photon pairs.  In particular,
areas outside of
the structure of function $\mathscr{L}(\textbf{k}_{s}^{\bot},
\textbf{k}_{i}^{\bot})$, which may be diagonal,  are ``removed'' from the plot of function
$|S(\textbf{k}_{s}^{\bot}+ \textbf{k}_{i}^{\bot})|^2$ and can yield a
narrowed and tilted CAS.

It is thus interesting to consider  the AS and CAS
in the short-crystal regime ($L \ll L_c$).  In this limit, the $\mathscr{L}(\textbf{k}_{s}^\bot,\textbf{k}_{i}^\bot)$ function
is much broader than the $|S(\textbf{k}_{s}^\bot+\textbf{k}_{i}^\bot)|^2$ function, so that the conditional angular spectrum is
determined by the latter, according to

\begin{eqnarray}\label{CASshort}
R^{(0)}_c(\textbf{k}^\bot_s,\textbf{k}_i^\bot)  \approx | S(\textbf{k}_{+}^\bot)|^2.
\end{eqnarray}

Note that because the CAS in Eq.~\ref{CASshort} depends only on the pump AS, it is azimuthally invariant.  It is also interesting to consider the amplitude underlying this conditional angular spectrum.   In the case of ideal idler detection involving a single transverse wavevector $\tilde{\textbf{k}}^\bot_i$ and frequency $\tilde{\omega}_i$, and within the short crystal regime, the state describing the heralded signal-mode single photon may be written as follows

\begin{equation}
|\Psi \rangle_s=\kappa \int d\textbf{k}^\bot_s A_s S(\textbf{k}^\bot_s+ \tilde{\textbf{k}}^\bot_i)|\omega_p-\tilde{\omega}_i,\textbf{k}^\bot_s\rangle.
\end{equation}

\noindent where $\kappa$ is a normalization constant.   It may be seen that under these conditions, the signal-mode single-photon wavevector amplitude constitutes a displaced version of the pump
wavevector amplitude, centered at $-\tilde{\textbf{k}}_i$.   Also, in the $L \ll L_c$ regime, $\mathscr{L}(\textbf{k}_{s}^\bot,\textbf{k}_{i}^\bot)$ is a slowly-varying function and may be
considered a constant for the purposes of the integration in Eq.~\ref{E:AS:PR}.  Thus, the AS is given as follows, where
we use the fact that the integral of the pump AS over all transverse wavevectors represents
the pump power, i.e. a constant,

\begin{eqnarray}
R^{(0)}_s(\textbf{k}^\bot_s)&\propto&  \mathscr{L}(\textbf{k}^\bot_s,-\textbf{k}^\bot_s).
\end{eqnarray}

\begin{figure}[ht]
\centering
\includegraphics[width=8cm]{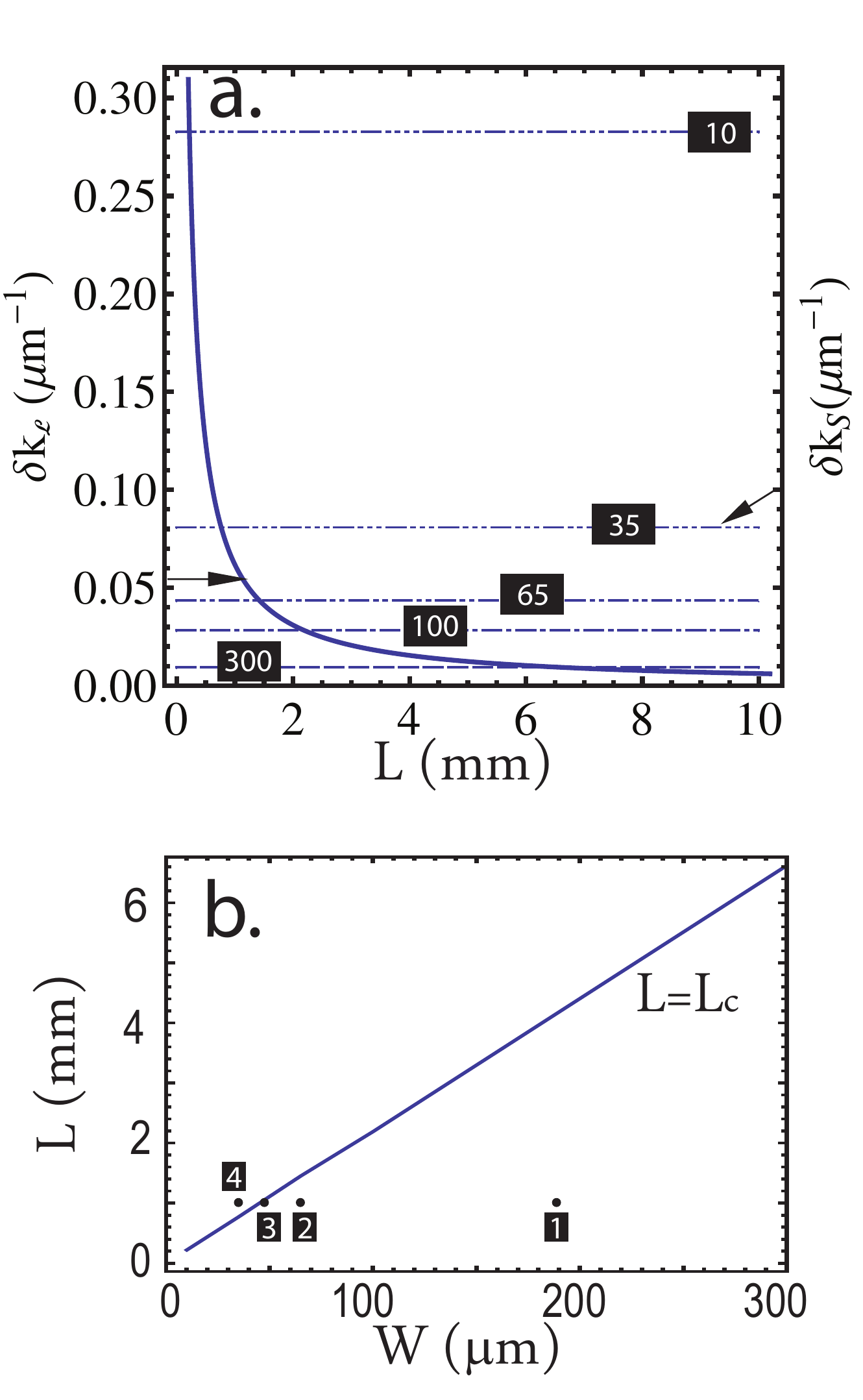}
\caption{(color online) In panel a) we show the widths of the functions $\mathscr{L}(\textbf{k}^\bot_s,\textbf{k}^\bot_i)$ and $|S(\textbf{k}^\bot_s+\textbf{k}^\bot_i)|^2$ as a function of the crystal length, for a fixed value of $\textbf{k}^\bot_i$ corresponding to the top of the AS, for different values of the pump beam radius indicated in microns within the black squares.  In panel b) we show the condition $L=L_c$ plotted in the parameter space $\{W,L\}$ obtained from the intersections in panel a.  We also indicate the four experimental measurements (see below) with labelled black dots.   \label{Fig:widths}}
\end{figure}

Thus, in the short-crystal regime, while the CAS depends only on the transverse phasematching
properties through the function $|S(\textbf{k}^\bot_{+})|^2$, the AS depends only on longitudinal phase matching
properties through the function $\mathscr{L}(\textbf{k}^\bot_s,-\textbf{k}^\bot_s)$.  In order to make the previous discussion more quantitative, let us define $\tilde{\textbf{k}}^\bot_i=(0,k^\bot_{iy})$ with $k^\bot_{iy}>0$ chosen so as to maximize the single-channel counts.  Then, we can define the $1/e$ full widths of the functions $|S(\textbf{k}^\bot_s+\tilde{\textbf{k}}^\bot_i)|^2$ and  $\mathscr{L}(\textbf{k}^\bot_s,\tilde{\textbf{k}}^\bot_i)$, along the $k_y$ direction,  as $\delta k_S$ and $\delta k_\mathscr{L}$, respectively.   Fig.~\ref{Fig:widths}(a) shows a plot of $\delta k_\mathscr{L}$ as a function of the crystal length obtained numerically (continuous line) for a BBO crystal with a $29.3^\circ$ cut angle; note that longer crystals lead to a smaller width $\delta k_\mathscr{L}$.
Fig.~\ref{Fig:widths}(a) also shows $\delta k_S$  for the case of a Gaussian beam pump, plotted with dashed lines for different values of $W=W_x=W_y$, (indicated, in microns, within the black rectangles).   We then define the critical crystal length $L_c$, for a given pump beam radius $W$, as that for which $\delta k_\mathscr{L}=\delta k_S$.

Fig.~\ref{Fig:widths}(b) represents the parameter space $\{W,L\}$, where we have assumed $W_x=W_y=W$, and where we include a plot of the condition $L=L_c$, obtained numerically, which turns out to have an essentially linear dependence on $W$; note that this condition cannot easily be obtained analytically.   This line divides
the parameter space in two parameter sub-spaces; the right-hand sub-space represents the set of all experimental configurations in the regime $L<L_c$, while the left-hand subspace represents the set of all experimental configurations in the regime $L>L_c$.  Also shown in the plot are four dots indicating our four experimental configurations (see discussion below; in particular dots 1 and 4 correspond to panels a and b of Fig.~\ref{Fig:fluxspdc}).  Thus, on the one hand, in the limit of a plane wave pump, $L_c \rightarrow \infty$, and the CAS is fully determined by the pump properties without any influence of the crystal properties regardless of the crystal length.  In this case, the CAS exhibits no variations around the SPDC annulus, and the photon pairs are thus azimuthally indistinguishable.   On the other hand, a greater degree of focusing (corresponding to smaller values of $W$), leads to a smaller critical crystal length $L_c$.  Thus, a sufficiently focused pump and/or a sufficiently long crystal implies that photon pairs are in the regime $L>L_c$, in which case the CAS becomes elongated and tilted leading to azimuthal distinguishability.

Related results were obtained in Ref.~\cite{burlakov97}.  In this paper, it was found  theoretically that for a sufficiently short crystal, and/or for sufficiently small emission angles,
the transverse variation of the crystal nonlinearity - pump amplitude product determines the conditional angular spectrum, while the angular spectrum is in this case determined solely by the nonlinear crystal properties.    Conversely, the paper by  Burlalkov {\it et al.} reports that for a sufficiently long crystal, and/or for sufficiently large emission angles, the nonlinear crystal properties determine the conditional angular spectrum while the angular spectrum becomes sensitive to the  transverse variation of the crystal nonlinearity - pump amplitude product.

\section{Experiment}

The objective of our experimental work presented here is to
characterize the angular distribution of the SPDC  photon
pairs, both in terms of single-channel and double-channel detection events where we
use variations in the degree of pump focusing to select whether the source is in the $L<L_c$ or
$L>L_c$ regime.

\begin{figure}[ht]
\centering
\includegraphics[width=8cm]{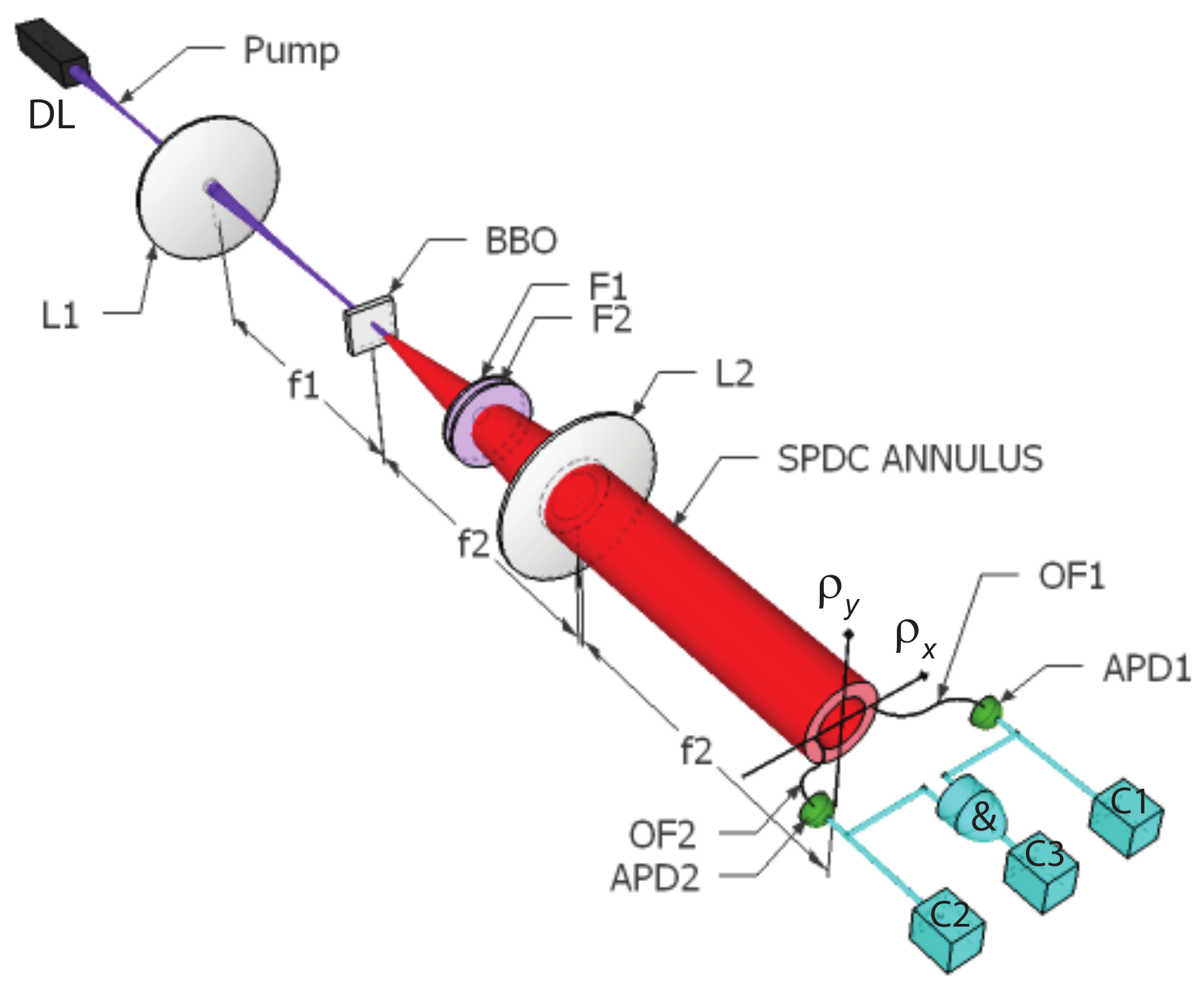}
\caption{(color online) Experimental setup used for measuring the angular spectrum,
and the conditional angular spectrum of SPDC photon
pairs.}\label{Fig:setup}
\end{figure}

Our experimental setup is shown schematically in
Fig.~\ref{Fig:setup}. A beam from a diode laser (DL) centered at
$406.8$nm is used as pump for the SPDC process. This beam is
spatially filtered by coupling into a single-mode fiber (not shown in the figure) and using the
collimated out-coupled beam, with $23$mW power. A blue colored glass
filter (Schott BG-39; not shown in the figure) is used in order to
suppress non-ultraviolet background photons.  The resulting beam
illuminates a $1$mm-thick $\beta$-barium borate (BBO) crystal, cut
at a phasematching angle of $29.3^\circ$ for type-I non-collinear
phasematching so that the degenerate photon pairs produced propagate
outside the crystal at an angle of $3.6^\circ$ with respect to the
axis defined by the pump beam.  Pump photons are suppressed by
transmitting the signal and idler modes through a long-pass filter
with a cut-on wavelength of $488$nm (F1), followed by a bandpass
filter centered at $810$nm with a $10$nm bandwidth (F2); both of
these filters are placed normal to the axis defined by the pump
beam.  A lens with a focal length of $10$cm (L2) is placed at a
distance of $10$cm from the crystal, thus defining the Fourier plane
a further $10$cm from the lens.

As discussed in the previous section, spatially-resolved photon
counting may be implemented with the help of spatial filters placed
on the Fourier plane,  leading to single-photon detectors (APD1 and
APD2).   In our experiment, we have used for this purpose the fiber tips of
large-diameter optical fibers (OF1 and OF2).   Note
that coupling of photon pairs into single-mode fibers is described
through the mathematical overlap between the two-photon state and
the fiber collection modes~\cite{vicent10}.  However, in the present
case where fibers are highly multi-mode, the incoherent sum of the
joint spectrum, projected onto all combinations of supported modes
makes detection phase-insensitive.   While in the case of the AS measurement a
single fiber tip is used, in the case of the CAS
measurement two separate fiber tips are used, one for each of the
signal and idler modes. The fiber tips are mounted so that they can
be displaced on the transverse plane, along the two perpendicular
directions: $x$, parallel to the optical table, and $y$ normal to
the optical table.  In the AS case, the fiber tip
displacement is carried out with computer-controlled linear motors
($50$nm resolution and $1.5$cm travel), and the fiber used has a
$200\mu$m diameter core. In the CAS case,
one of the fiber tips (corresponding to the idler mode) can be
translated manually along the two axes, while the other fiber tip
can be translated with our computer-controlled linear motors. Both
of the fibers used have a $200\mu$m diameter core.

For the AS measurement, the fiber tip scans a
sufficient transverse area in order to encompass the entire emission
annulus.  For the CAS measurement, a
location $\textbf{k}^\bot_{i0}$ for the idler-mode fiber tip is
selected on the SPDC annulus, which determines by transverse
momentum conservation the expected location, $-\textbf{k}^\bot_{i0}$,
for the conjugate signal photons.  In our experiments we have chosen
$\textbf{k}^\bot_{i0}$, with a vanishing $y$ component
$k_{iy0}$, and with a negative value of $k_{ix0}$
(left side of the SPDC cone looking into the crystal) chosen so that
the number of counts is maximized.   The signal-mode fiber tip is
then scanned over an area around  $-\textbf{k}^\bot_{i0}$.

The optical fibers (a single one for the AS
measurement, and two of them for the CAS
measurement) lead to fiber-coupled silicon single-photon counting
modules (SPCM's).  The electronic pulses generated by the SPCM's are
inverted, attenuated and discriminated to produce standard nuclear
instrumentation module (NIM) pulses of 7ns duration.   These signals
are on the one hand directly counted with pulse counters (C1 and C2
in Fig.~\ref{Fig:setup}), to yield single-channel counts. On the other
hand, these signals form the inputs for an AND gate (\&) which
produces an output pulse when the two inputs are temporally
overlapped. The output from the AND  gate is counted by a third
pulse counter (C3), to obtain the coincidence counts. We have in the
region of $200$ background counts per second, including dark counts,
in each of our two detectors.  This level of background counts leads
to essentially no accidental coincidence counts related to dark
counts.

We have carried out AS and CAS measurements for four different
pump-beam focusing strengths.   These situations correspond to: i)
no focusing lens used, and to a focusing lens with the following
focal lengths used: ii) $f_1=30$cm, iii) $f_1=10$cm, and iv)
$f_1=6$cm. In all cases, the lens is placed a distance of one focal
length from the crystal.  Note that the resulting pump beamwaist
is not necessarily precisely centered with respect to the crystal;  while in our
theory, the AS and CAS depend
on the beam radii $W_x$ and $W_y$ at the beamwaist, these functions do not depend on the beamwaist location with respect to the  crystal's center
plane~\cite{vicent10}. The values of $W_x$ and $W_y$, directly
measured by recording the beam profile with a CCD camera at a number
of distinct  propagation planes and fitting to the standard beam
radius vs propagation distance expression for Gaussian beams, are
shown in Table~\ref{Table}, along with the resulting critical length $L_c$.  Note that since
the crystal length used is $1$mm, measurements 1 and 2 are in the $L<L_c$ regime, while
measurement 3 is essentially on the boundary and measurement 4 is in the
$L>L_c$ regime.  Note also that the choice of parameters for each of the four measurements is indicated in Fig.~\ref{Fig:widths}(b) by
labelled dots, where the horizontal coordinate is determined by the corresponding $W_y$ value from Table ~\ref{Table}; indeed, for a fixed
detector at the top of the AS, the CAS depends largely on $W_y$.   The values of  $W_x$ and $W_y$ shown in the table were
used for the numerical simulations of the AS and of
the CAS to be presented below for each of
the measurements (1 through 4).

\begin{table}[h]
\centering
\begin{tabular}{|c|c|c|c|cl}
\hline
Measurement             & $W_{x}$ ($\mu$m)          & $W_{y}$ ($\mu$m)  & $L_c$ (mm) \\
\hline
1) no lens used         & 182.0                       & 189.0   & 4.1\\
2) $f_1=30$cm             & 67.5                      & 64.8  & 1.4 \\
3) $f_1=10$cm             & 56.4                      & 47.9 & 1.1\\
4) $f_1=6$cm              & 38.9                     & 34.7 & 0.8 \\
\hline
\end{tabular}
\caption{Vertical ($W_y$) and horizontal ($W_x$) beam widths, at the
beamwaist, measured for each of the four measurements.}\label{Table}
\end{table}

\begin{figure*}[ht]
\centering
\includegraphics[width=11cm]{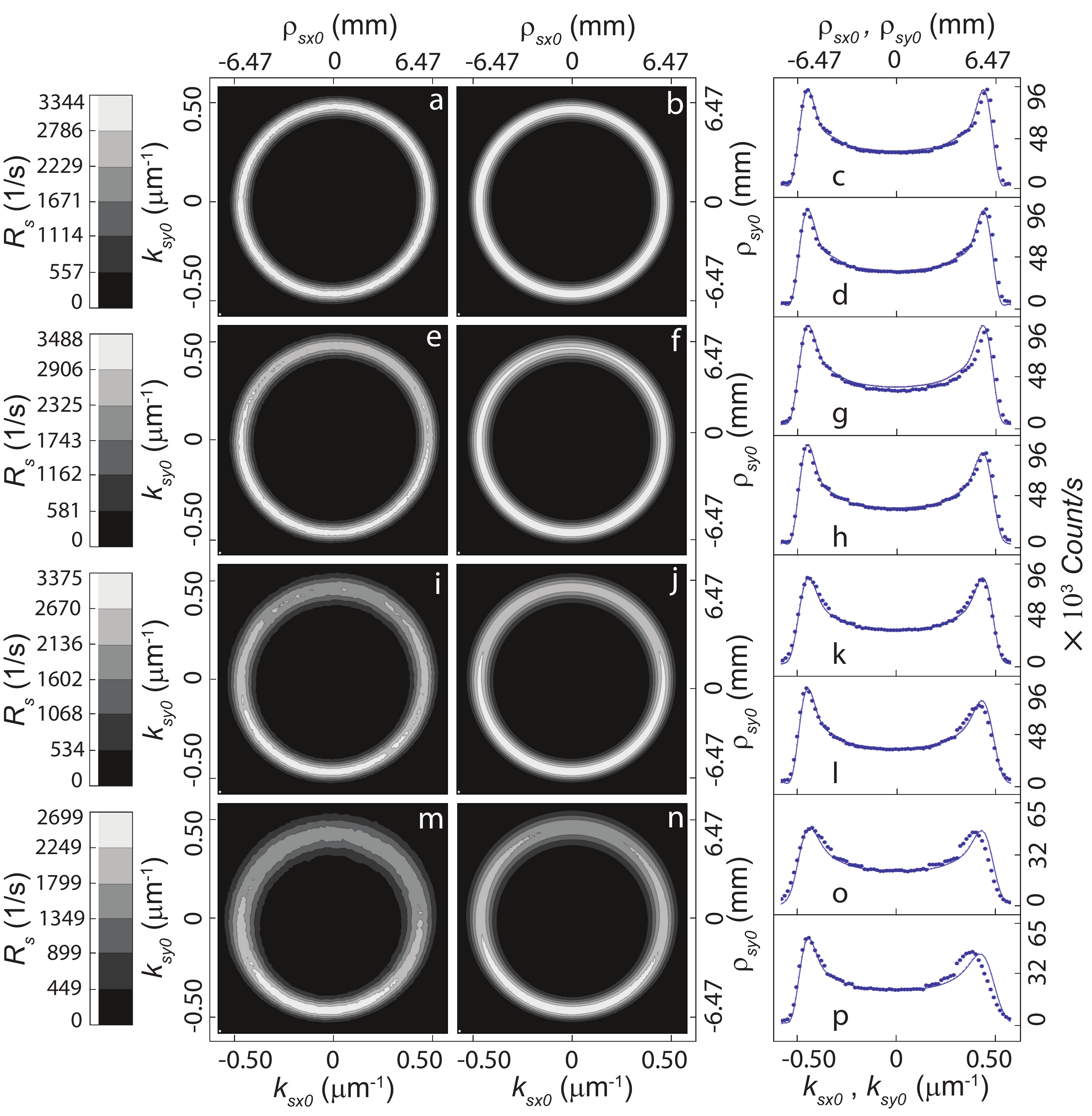}
\caption{For measurement 1: (a) Experimentally-measured angular
spectrum, with level of counts shown in gray-scale bar, (b)
Corresponding numerical simulation, (c) The dots show the result of
adding up the rows of the matrix of values in panel (a), the line
shows the AS integrated over the $\rho_{sx0}$
coordinate, (d) The dots show the result of adding up the columns of
the matrix of values in panel (a), the line shows the numerically-obtained angular
spectrum integrated over the $\rho_{sy0}$ coordinate.    Blocks of
panels (e)-(h), (i)-(l), and (m)-(p) are similar to block (a)-(d)
for each of measurements $2$,$3$ and $4$. } \label{Fig:single-channel}
\end{figure*}

For each of these four cases, we have carried out a measurement of
the AS, and a corresponding numerical simulation.
These results are shown in Figure~\ref{Fig:single-channel}, which is
organized in four blocks, for each of the focusing strengths from
Table~\ref{Table}. Panels (a)-(d) correspond to measurement $1$,
panels (e)-(h) to measurement $2$, and so forth.  Within each of
these blocks, the first panel represents a measurement of the
AS shown in six gray levels, as indicated by the gray
level bar on the left.  Note that background counts have been
subtracted for each of the AS measurements.    For
these measurements, data was taken on a transverse position grid,
involving a counting period of $1$ second at each point.  Each grid
point represents a particular transverse position $\bm{\rho}_{s0}^\bot$ of the fiber tip, which corresponds to a
transverse momentum value $\textbf{k}_{s0}^\bot=[\omega_s/(cf_2)]
\bm{\rho}_{s0}^\bot$. A grid spacing of $200\mu$m is used for
measurements $1-3$ and of $250\mu$m is used for measurement $4$ in
AS measurements, while a larger spacing of $400\mu$m
is used for measurements $1-3$ and of $500\mu$m for measurement $4$
in areas of low counts, e.g. inside the annuli. The second panel
represents the corresponding numerical simulation, where we have
scaled the maximum number of counts to coincide with the
experimentally-obtained maximum number of counts.  The specific
simulation carried out yields $R_s(\bm{\rho}_{s0}^\bot)$ by
numerical integration of Eq.~\ref{E:AS:PR}; for convenience, we
have also labeled these plots with the transverse momentum values at
the degenerate SPDC frequency.  Note that because the transverse
dimensions of the fiber used for photon collection is negligible
compared with the width of the AS annulus, in
computing our numerical simulations we have used the expression
corresponding to delta-like detectors on the Fourier plane
(Eq.~\ref{E:AS:PR} rather than Eq.~\ref{E:AS}).  

The third and fourth panels in each block show projected angular
spectra obtained by adding together values along columns of this
grid, to obtain the horizontal projected AS, and
likewise obtained by adding together values along rows of this grid
in order to obtain the vertical projected AS.  We
employ these projected angular spectra for a careful comparison
between measurements and simulations; note that while we could also
use for this purpose a ``slice'' obtained for fixed $\rho_{sx0}$
(or fixed $\rho_{sy0}$), the projected angular spectra lead to
considerably more counts per grid location, and therefore to better
statistics. The continuous lines represent the corresponding
numerical simulations, where the AS has been
integrated over the $\rho_{sy0}$ coordinate to obtain the horizontal
projected AS, and over the $\rho_{sx0}$ coordinate to
obtain the vertical projected AS.  In general terms it
may be seen that increasing the degree of pump focusing (or
decreasing $W_x$ and $W_y$) leads to an increasingly asymmetrical
AS, along the vertical direction, with a larger width
at the top of the annulus than at its bottom;  note that, in
contrast, the widths at the left and right of the annulus are
comparable.   As discussed above, this AS asymmetry appears for parameter
combinations in the regime $L>L_c$.  This asymmetry, which is related to Poynting vector
walkoff, is clear from the vertical projected AS which
exhibits a wider and shorter right-hand peak compared to the
left-hand peak. In contrast, the horizontal projected angular
spectrum is symmetric, both peaks exhibiting identical heights and
widths. Related results obtained with a CCD camera have been
reported, for type-II SPDC,  in Refs.~\cite{lee05,bennink06} and
using a LED pump in Ref.~\cite{tamosauskas10}.  Note that the
agreement between the experimental measurements and the numerical
simulations is excellent.

\begin{figure*}[ht]
\centering
\includegraphics[width=11cm]{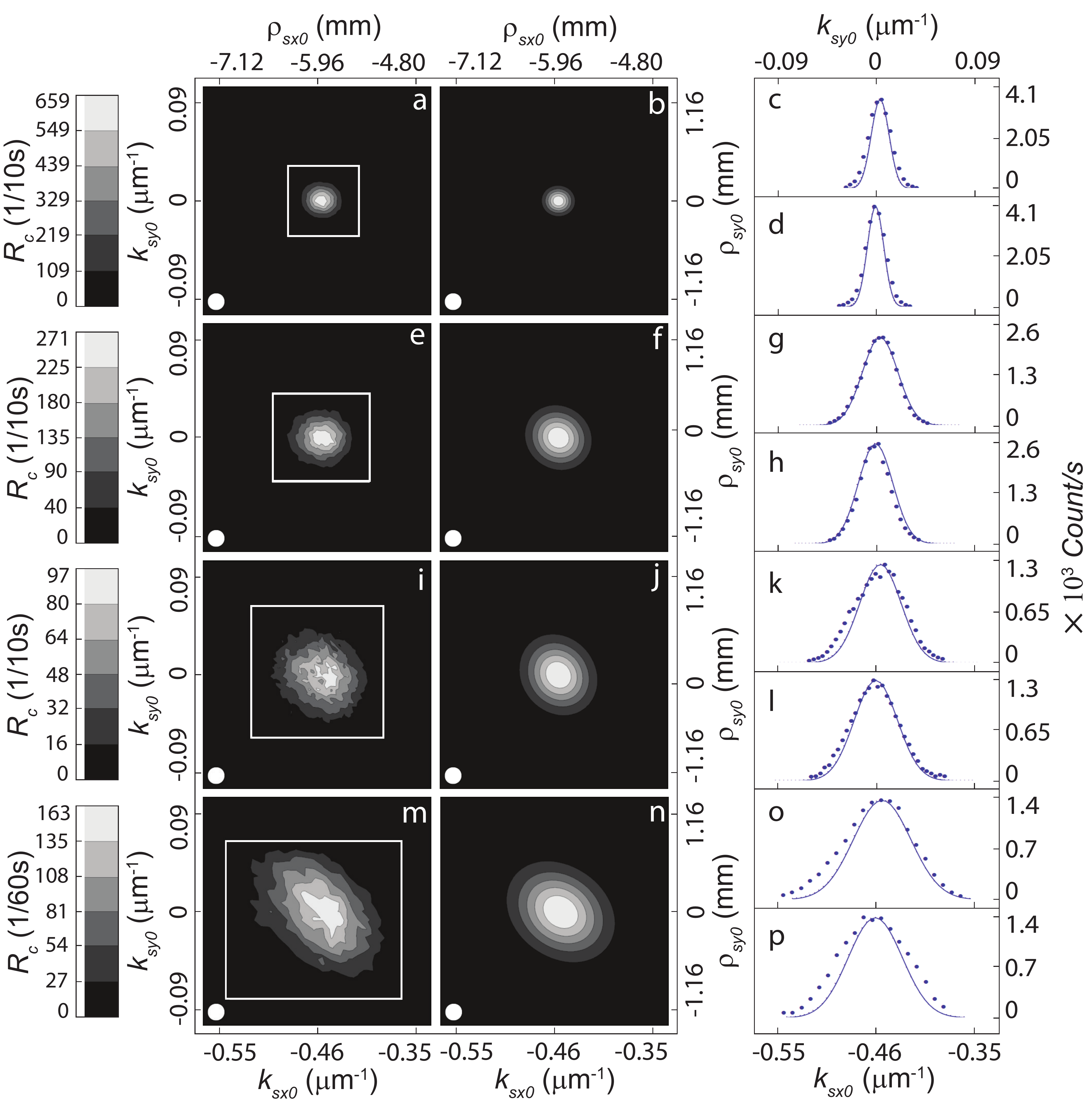}
\caption{For measurement $1$: (a) Experimentally-measured
conditional angular spectrum, with level of counts shown in
gray-scale bar, (b) Corresponding numerical simulation, (c) The dots
show the result of adding up the rows of the matrix of values in
panel (a), the line shows the conditional angular spectrum
integrated over the $\rho_{sx0}$ coordinate, (d) The dots show the
result of adding up the columns of the matrix of values in panel
(a), the line shows the numerically-obtained conditional angular spectrum integrated over
the $\rho_{sy0}$ coordinate.   Blocks of panels (e)-(h), (i)-(l),
and (m)-(p) are similar to block (a)-(d) for each of measurements
$2$,$3$ and $4$.}\label{Fig:doubles}
\end{figure*}

Let us now turn our attention to coincidence counts, i.e. to
measurements of the \textit{conditional} angular spectrum. As in the
case of our measurements of the AS, we have undertaken
measurements for the four experimental situations from
Table~\ref{Table}.  Likewise, for each of these four cases we have
carried out a corresponding numerical simulation.  These results are
shown in Figure~\ref{Fig:doubles}, which is organized in four
blocks, for each of the focusing strengths from Table~\ref{Table}.
Panels (a)-(d) correspond to measurement $1$, panels (e)-(h) to
measurement $2$, and so forth.  Within each of these blocks, the
first panel represents a measurement of the conditional angular
spectrum, shown in six gray levels, as indicated by the gray level
bar on the left.    For these measurements, data was taken on a
transverse position grid, located around the transverse position
conjugate to the position of the idler-mode fiber. Each grid point
represents a particular transverse position $\bm{\rho}_{s0}^\bot$
of the fiber tip, which corresponds for SPDC frequency $\omega_s$ to a
transverse momentum value $\textbf{k}_{s0}^\bot=[\omega_s/(cf_2)]
\bm{\rho}_{s0}^\bot$. We have used  a counting period of $10$
seconds at each point for measurements $1-3$, and of $60$ seconds at
each point for measurement $4$.  These counting periods reflect the
fact that for a greater degree of focusing, the counts become spread
out over a greater transverse area, so that the level of counts at
each grid point is reduced. A grid spacing of $50\mu$m is used for
measurements $1-3$ and of $100\mu$m is used for measurement $4$. The
white frame which encompasses the region with counts represents the
range of transverse positions where data was taken; for positions
outside of this frame, the coincidence counts were fixed to zero in
the plots.  We show the transverse dimensions of the fiber core used
for photon collection through a white disk appearing near the
bottom-left corner of each panel.  It may be appreciated that the
transverse extent of the collection fiber can be significant
compared to the width of the measured CAS.
In contrast, note that in the case of the AS (single-channel
counts; Fig~\ref{Fig:single-channel}), the transverse dimensions of the
fiber may be neglected, since they are much smaller than the width
of the AS annulus.  In fact, a white disk which
represents the fiber core transverse dimensions is shown in
Fig.~\ref{Fig:single-channel}, although it is difficult to see due to its
small size.

The second panel in each block represents the corresponding
numerical simulation, where we have scaled the maximum number of
counts to coincide with the experimentally-obtained maximum number
of counts.  Note that for these simulations we have assumed that the
spatial filter functions $u_s(\textbf{k}^\bot_s-\textbf{k}^{\bot}_{s0})$
and $u_i(\textbf{k}^\bot_i-\textbf{k}^{\bot}_{i0})$ are Gaussian with a
full width at $1/e$ of $200\mu$m.  The specific simulation carried
out yields $R_c(\bm{\rho}_{s}^\bot, \bm{\rho}_{i0}^\bot)$ by numerical integration of a
version of Eq.~\ref{E:CAS} written in terms of transverse
position; for convenience, we have also labeled
these plots with the transverse momentum values at the degenerate
SPDC frequency.  Note that because in the case of the CAS the transverse width of the fiber is significant, in
computing our numerical simulations, we have used
Eq.~\ref{E:CAS}, which takes into account the transverse extent
of detectors on the Fourier plane, rather than Eq.~\ref{E:CAS:PR}
which assumes delta-like detectors. Note also that in the case of
coincidence counts, there are no background counts; the plots show
the actual number of counts without subtracting a background level.
The maximum number of coincidence counts decreases as the strength
of focusing is increased so that the data is of greater quality for
lower focusing strengths.    The third and fourth panels in each
block show projected conditional angular spectra obtained by adding
together values along columns of this grid, to obtain the horizontal
projected AS, and likewise obtained by adding together
values along rows of this grid in order to obtain the vertical
projected AS.  As in the case of single-channel counts we
employ these projected angular spectra for a careful comparison
between measurements and simulations.  The continuous lines
represent the corresponding numerical simulations, where the angular
spectrum has been integrated over the $\rho_{sy0}$ coordinate to
obtain the horizontal projected AS, and over the
$\rho_{sx0}$ coordinate to obtain the vertical projected AS. As can be appreciated, the agreement is excellent.

It is clear from our experimental and numerical results that an increased level of pump focusing
broadens the CAS, and that for a sufficiently long crystal ($L>L_c$), the CAS may become tilted.   Indeed,
as expected from our theory, the CAS in fact corresponds to a displaced pump angular spectrum, which for $L>L_c$ may become clipped by
function $\mathscr{L}(\textbf{k}^\bot_s,\textbf{k}^\bot_i)$ and can then become tilted.  Thus, a greater degree of pump focusing leads to a broader pump angular spectrum, and this in turn leads to a broader CAS.

Note that for, both, the AS and CAS measurements, in the case of measurement $4$, i.e. the most
highly focused case that we have considered, the agreement is not as
optimal as for measurements $1-3$.   We have observed that despite
the use of spatial filtering through a single mode fiber, for an
increasing degree of focusing, the pump beam acquires additional
structure and becomes progressively less Gaussian.  Since our theory
assumes a perfectly Gaussian pump beam, this explains the observed slight
discrepancy between theory and experiment for measurement 4.

\section{Conclusions}

We have presented a theoretical and experimental exploration of the joint effects of the
pump transverse electric field distribution and of the non-linear crystal on the
properties of photon pairs generated by spontaneous parametric downcinversion (SPDC).  We have
focused this analysis on the angular spectrum (AS) and on the conditional angular spectrum (CAS) of the SPDC photon pairs.
We have shown
that the CAS may be written as the product of two functions, one which
is related to transverse phasematching and depends
on pump properties, and another one which is related to longitudinal phasematching and depends on nonlinear crystal properties.
We have shown that a critical crystal length $L_c$ exists, which depends on the degree of pump focusing,
such that for $L<L_c$ the CAS is fully determined by the pump AS, and that for
$L>L_c$ the CAS is determined jointly by crystal and pump properties.  For a Gaussian
beam pump,  $L_c$ turns out to have an essentially  linear relationship with the beam radius.    Thus, the condition $L=L_c$ divides the $\{W,L\}$ parameter
space into two separate parameter subspaces, where $L<L_c$ leads to a symmetric AS and to an
azimuthally invariant CAS, and where $L>L_c$ leads to an asymmetric AS and to a CAS which
varies in width and orientation around the SPDC annulus.

We have also presented experimental measurements of the AS and CAS for photon pairs generated
through type-I non-collinear spontaneous parametric downconversion.
These measurements were carried out by spatially-resolved photon
counting, and by coincidence spatially-resolved photon counting,
respectively.    We have presented experimental data for the AS and CAS, along with corresponding
numerical simulations  based on our theory, for four different experimental configurations amongst which the degree of pump focusing is varied. A comparison of our experimental measurements with our numerical simulations leads to excellent agreement.
Of the four experimental configurations used, two are in the regime $L<L_c$, one is near  $L=L_c$, and one is in the regime
$L>L_c$.     Our measurements show that, as expected from our theory,  pump focusing leads to an
asymmetric broadening of the AS, and to a broadening and tilting of
the CAS. Physically, the broadened AS is a
consequence of the greater
spread of pump transverse wavevectors, resulting in  phasematching for a greater
spread of signal and idler transverse wavevectors.  This results in broadening of the conditional angular
spectrum,  so that
each idler-mode
$k$-vector is correlated to a spread of signal-mode $k$-vectors,
while this correlation is one-to-one in the idealized case of a
plane-wave pump.     We believe that these results will lead to an
enhanced quantitative and qualitative understanding of the spatial properties of
type-I, non-collinear spontaneous parametric downconversion photon
pairs and to an important tool for source design.

\begin{acknowledgements}
This work was supported in part by CONACYT, Mexico,  by DGAPA, UNAM
and by FONCICYT project 94142.
\end{acknowledgements}

\end{document}